\pdfoutput=1
\documentclass{bmcart}

\usepackage[utf8]{inputenc} 
\usepackage{soul,color}
\usepackage{subfig}
\usepackage{todonotes}
\usepackage{graphicx}
\graphicspath{{./figures/}}
\DeclareGraphicsExtensions{.pdf,.jpeg,.png,.jpg}

\startlocaldefs
\endlocaldefs

\begin{document}

\begin{frontmatter}

\begin{fmbox}
\dochead{Research}


\title{Challenges in Mobile Multi-Device Ecosystems}


\author[
   addressref={aff1},                   
   email={jg@jensgrubert.de}   
]{\inits{JG}\fnm{Jens} \snm{Grubert}}
\author[
   addressref={aff1},                   
   email={matthias.kranz@uni-passau.de}   
]{\inits{MK}\fnm{Matthias} \snm{Kranz}}
\author[
   addressref={aff2},                   
   email={aquigley@st-andrews.ac.uk}   
]{\inits{AQ}\fnm{Aaron} \snm{Quigley}}

\address[id=aff1]{
  \orgname{University of Passau}
}
\address[id=aff2]{%
  \orgname{St Andrews University},
}


\begin{artnotes}
\end{artnotes}

\end{fmbox}


\begin{abstractbox}

\begin{abstract} 
Coordinated multi-display environments from the desktop, second-screen to gigapixel display walls are increasingly common. Personal and intimate mobile and wearable devices such as head-mounted displays, smartwatches, smartphones and tablets are rarely part of such multi-device ecosystems. With this paper, we contribute to a better understanding about factors that impede the creation and use of such mobile multi-device ecosystems.  We base our findings on literature research and an expert survey. 
Specifically, we present grounded challenges relevant for the design, development and use of mobile multi-device environments.\end{abstract}




\end{abstractbox}
%

\end{frontmatter}





\section{Introduction}
\label{intro}

\begin{figure}[th]\center
\includegraphics[width=\textwidth]{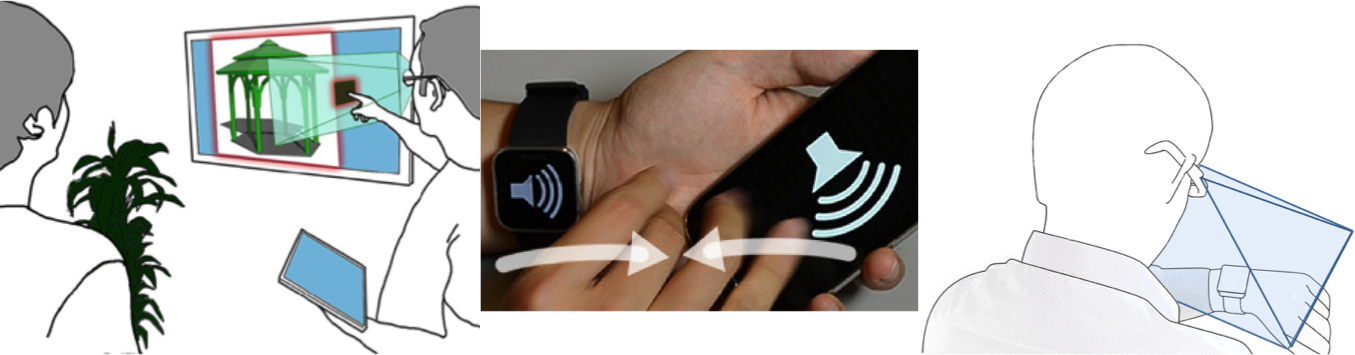}
    \caption{Mobile multi-device environments. Left: interaction between head-mounted display, tablet and public display (image courtesy of Serrano et al.) \cite{Serrano:2015:GDH:2785830.2785838}. Middle: interaction between a smartphone and smartwatch (image courtesy of Chen et al.) \cite{chen2014duet}. Right: Interaction between head-mounted display and smartphone \cite{grubert2015multifi}.}
    \label{fig:teaser}       
\end{figure}

Multi-display environments from the desktop to gigapixel displays have emerged as ubiquitous interfaces~\cite{Krumm:2009} for knowledge work (e.g., Microsoft Surface Hub for collaboration or Bloomberg systems for financial trading) and complex tasks (e.g. city or factory management). Similarly, social applications such as second screen TV experiences are further extending the proliferation of increasingly complex display ecosystems with different sizes, mobility or reachability. In parallel, we see the emergence of further classes of more personal, intimate and body-centric computing in the form of head-mounted displays (HMDs) such as the Oculus Rift or Microsoft Hololens and smartwatches such as AndroidWear or Apple Watch, which promise always-on information access around the user’s body. Small touch devices (such as smartwatches and smartphones) aim at improving mobility, portability and privacy by simply shrinking the device, but as a result sacrifice the display and interaction area. HMDs have the potential to enable rich spatial interaction with information located around the user’s body through dexterous and expressive human hand motion, but at the cost of interaction accuracy, and, like wearables, present challenges for sharing information with co-located people. Support for activities between and across a set of devices around the user’s body presents a myriad of challenges, which we aim to address in this paper. 


\emph{Mobile multi-device environments} promise to overcome limitations of interacting with individual devices in mobile contexts. These environments consist of multiple interactive and \emph{coordinated devices}, typically displays, with at least one mobile or wearable component (see Figure \ref{fig:teaser}). We see recent interest in the research community (e.g., \cite{Serrano:2015:GDH:2785830.2785838, chen2014duet, grubert2015multifi, houben2015watch}).  However, while, for example, multi-display interaction is already common in stationary scenarios, to date we see less cross-device and multi-display interaction \cite{santosa2013field, jokela2015diary}, which include mobile or wearable components employed outside of the laboratory. We argue that in order to integrate such diverse and disparate types of displays into a unifying interaction environment and to enable interaction with information freely across device boundaries, we need to better understand specific challenges for the creation and use of such systems. With this article we aim to contribute an overview on challenges for designing, developing, evaluating and using mobile multi-device environments. We base our findings on literature research and reflections about the development of mobile multi-device environments by the authors. Furthermore, we complement these findings with the results and analysis of an expert survey.



\section{Challenges in Mobile Multi-Device Environments}
The fundamental challenges in mobile multi-device ecosystems reach beyond that of multi-device ecosystems~\cite{Terrenghi:2009}. This is connected to the larger variety of input and output modalities found on mobile devices compared to desktop systems, to the mobility of each individual component and to the proximity of those devices to the human body. We aim at uncovering relevant challenges that impede the creation and use of such systems.


\subsection{Methodology}
\begin{figure}[t]\center
\includegraphics[width=\textwidth]{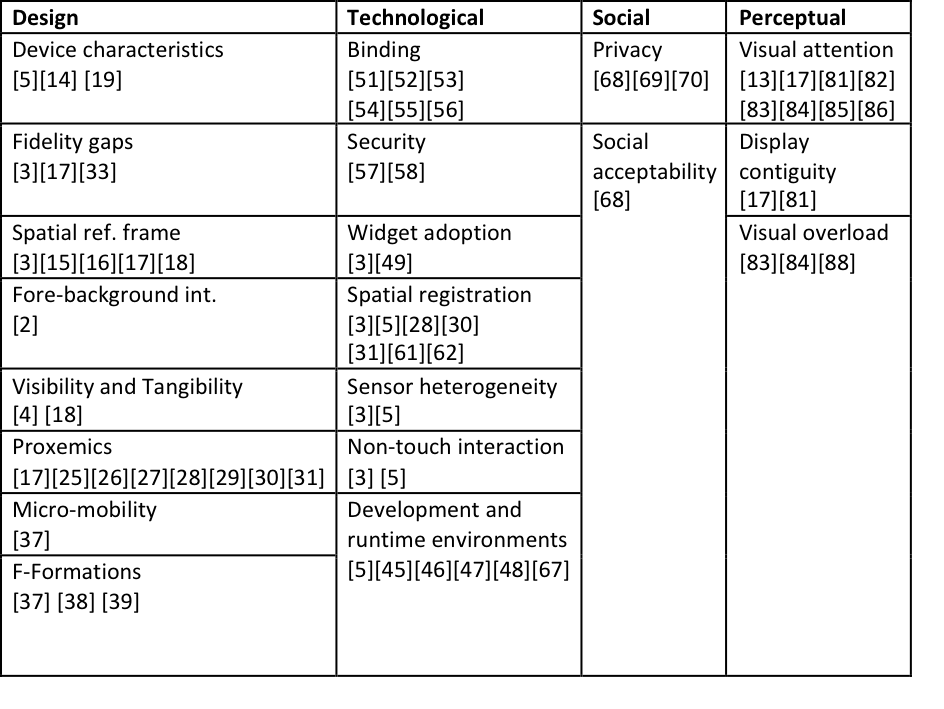}
    \caption{Challenges mentioned in papers in the domain of multi-device ecosystems.}
    \label{fig:cat}       
\end{figure}

To identify the key relevant challenges, we developed a review protocol for conducting our literature survey to ensure maximum coverage of relevant publications. This section describes our review protocol, starting with a literature survey using the ACM digital library.
The ACM digital library keyword search returned 37 entries for the keyword ``multi-display interaction'', 94 papers for ``multi-device interaction'', 94 for ``cross-device interaction'', 0 for ``cross-surface interaction'', 10 for ``cross-display interaction'', 5 for ``multi-fidelity interaction'', 30 for ``distributed displays'', 142 for ``distributed user interfaces'', 16 for ``second-screen interaction'', and 6 for ``multi-screen interaction’’, in total 434 prior to de-duplication. To minimise the possibility of excluding relevant research, relevant proceedings (e.g., ACM CHI, MobileHCI, UIST, AVI, DIS, EICS, ITS/ISS, INTERACT, HCI International, NordiCHI) and journals (e.g., TOCHI) were then searched for papers with explicitly identified challenges relevant for mobile scenarios. In addition, given this initial set of papers, we extended our search to secondary papers mentioned in the reference sections or referring to the initial papers (via Google Scholar). Furthermore, we scrutinised more closely without the above search criteria, the papers from recent workshops in this domain \cite{schneegass2015mobile, luceromobile, houben2015cross}. In total, we considered 140 papers (excluding papers which where redundantly identified through the keyword search) by matching their titles and abstracts. Starting from a broad set of keywords (standard codes) we created classifications through open and axial coding steps \cite{strauss1990basics}. With axial coding we employed our keywords to identify central ideas, events, and usage conditions and strategies which we grouped into categories and sub-categories. While open coding allowed us to identify new concepts and join them into categories and sub-categories. The merger of the classifications between open and axial coding produced the categories and sub categories we detail in the following sections. In addition to a literature survey, we also reflected on our own experiences in researching multi-device environments \cite{grubert2015multifi, Terrenghi:2009, rashid2012cost}. 
In the subsequent sections, we have grouped challenges into four top-level categories of design, technological/development, social and perceptual/physiological challenges. However, we are aware that some challenges can be associated with multiple categories (e.g., device-binding can be seen from a technical development or user-centred design point of view). Specific aspects of challenges associated for multiple top-level categories are discussed in the relevant sections. The results of the individual sections are summarized in Figure \ref{fig:cat}. In order to be associated with a category, a paper had to explicitly mention relevant aspects of that category.

\subsection{Design Challenges}

There are a number of design challenges for realizing mobile multi-device ecosystems for single user and collocated interaction.
For single user interaction these challenges include varying device characteristics, fidelity, spatial reference frame, foreground-background interaction, visibility and tangibility. For collocated interaction, we additionally identified micro-mobility, f-formations, and space syntax. Several design factors that are potentially relevant for mobile multi-device interactions, have been identified in previous work. In total 26 papers fall into this category which we divided into nine subcategories.  

\emph{Parameterization} i.e. characteristics of individual devices, e.g.,  ID, pose, data context and (prior) selection on the phone or smartwatch has been explored by Schmidt et al. \cite{schmidt2012cross} as well as Houben et al. \cite{houben2015watch} to describe how the interaction on a large interactive surface could be supported. Similarly, Grubert et al. used the term \emph{fidelity} to describe the quality of output and input characteristics, such as resolution, colour contrast, fixed vs. variable focus distance of devices in a mobile multi-device system \cite{grubert2015multifi}.

\emph{Spatial Reference Frame}; i.e. the real-world entity, relative to which interaction takes place is explored in terms of the roles adopted in several papers~\cite{grubert2015multifi, Hinckley:1994:SDI:192426.192501, billinghurst1999wearable, Cauchard:2011:VSM:2047196.2047256, ens2014ethereal}. Examples include body-parts of the user (head, chest, hands), physical objects in the scene (table, monitor, mug, poster, other mobile devices) or world-referenced locations (longitude and latitude).

\emph{Pairwise device interaction} has also explored how two touch screens could be used together by \emph{enabling or disabling their input and output channels}, including combinations of smartphones with (large) interactive surfaces \cite{serrano2015sw}, smartwatches with interactive surfaces \cite{houben2015watch}, or smartglasses with smartwatches \cite{serrano2015sw}, resulting in four different device combinations.  

\emph{Foreground-background interaction} \cite{buxton1995integrating, hinckley2005foreground} was applied to mobile multi-device environments by Chen et al. \cite{chen2014duet}. Foreground activities require attention (e.g., dialing a number); they are intentional activities. Background activities take place in the periphery, requiring less attention (e.g., being aware of a nearby person). Ideally, background activities can be sensed and actions can be triggered automatically (e.g., automatically switching on the light when a person enters a room). Chen et al. explored interaction techniques when both a smartphone and a smartwatch were jointly used as foreground devices \cite{chen2014duet}.

\emph{Proxemic dimensions} \cite{hall1969hidden, greenberg2011proxemic, marquardt2015proxemic} have also been applied to mobile multi-device scenarios (e.g., \cite{Cauchard:2011:VSM:2047196.2047256, Marquardt:2012:GEF:2396636.2396642, Hamilton:2014:CEU:2556288.2557170, Radle:2015:SSE:2702123.2702287, radle2014huddlelamp, Sorensen:2012:ISM:2399016.2399094, Lucero:2011:PCU:1978942.1979201, Lucero:2012:MCU:2371574.2371634}. Proxemics can be understood as culturally dependent ways in which people use interpersonal distance to understand and mediate their interactions with other people. Greenberg et al. identified distance, orientation, movement, identity, and location as relevant proxemic dimensions for ubiquitous computing \cite{greenberg2011proxemic}. More recently, Proxemics can be seen as a form of context-awareness for supporting users' explicit and implicit interactions in a range of uses, including remote office collaboration, home entertainment, and games~\cite{greenberg_et_al:DR:2014:4436}. Beyond such simple proxemics we suggest the need to consider kinesics, paralinguistics, haptics, chronemics and artifacts around us in our understanding of the design challenges. 

There are a number of further design factors, which have not yet been explored in depth. For example, Grubert et al. presented \emph{continuity of fidelity / fidelity gaps} as a relevant design factor. Continuity of fidelity can be understood as the degree to which individual device characteristics differ across devices, specifically input modalities (e.g., touch vs. in-air gestures or input resolution) and output modalities (such as display size, resolution, contrast). One need only consider the fidelity of inputs possible with a Microsoft Kinect, Leap Motion, Touch Screen or Google’s Project Soli or the size and display resolution on a Microsoft Surface Hub, Microsoft Band, or smartwatch to appreciate the challenge continuity of fidelity presents. Cauchard identified similar challenges \cite{Cauchard:2011:VSM:2047196.2047256, cauchard2013towards}. Ens et al. identified a number of design factors focused on interaction with 2D information spaces \cite{ens2014ethereal}. While not directly targeted at multi-device use, some of these factors appear to be relevant. For example, \emph{tangibility} describes if the presented information is perceptible by touch~\cite{Krumm:2009}. For example, touch screens provide a tangible representation of information spaces with haptic feedback. Virtual screens in optical see-through head-mounted displays such as Google Glass or Microsoft HoloLens or projectors are typically not tangible. Very recent work on mid-air haptic feedback using ultrasound promises to add tangibility even for those projection-based displays \cite{Long:2014:RVH:2661229.2661257, marzo2015holographic}. Another relevant design dimension is the \emph{visibility} of the individual devices and information spaces; i.e. the amount of visual information available in a multi-device interface \cite{ens2014ethereal}. The visibility also determines the degree to which proprioception is needed for operating an interface.

\emph{Co-located Interaction} in mobile multi-user, multi-device scenarios present additional factors we can identify. For example, \emph{micro-mobility} is the fine-grained positioning and orientation of objects so that those objects might be fully viewed, partially viewed or hidden from other persons \cite{luff1998mobility, marquardt2012cross}. \emph{F-formations} are spatial patterns formed during face-to-face interactions between two or more people \cite{marquardt2012cross, kendon1990conducting,  Marshall:2011:UFA:1958824.1958893}. Another potentially relevant design framework is \emph{space syntax} \cite{hillier2007space, schroder2007quantifying}. Originally aimed at urban planning, space syntax is ``a family of techniques for representing and analysing spatial layout of all kinds'' \cite{hillier2007space}.


However, to date it remains unclear if the described design factors are sufficient for guiding future design space explorations, if and how they are interdependent, to which extent they are relevant for non-touch screen devices and how they scale to more than two jointly used displays. For example, fidelity gaps might be more relevant for touch-screen - smartglass interaction as the difference in output resolution and contrast is considerably larger compared to interaction with two touch screens only \cite{grubert2015multifi}. Further challenges for the interaction design of multiple wearable displays concern how to explicitly or implicitly transition between individual interaction modes, e.g., from side-by-side to device-aligned \cite{grubert2015multifi}, from touch to mid-air interaction \cite{marquardt2011continuous, chen2014air+} and viewing \cite{Serrano:2014:ISP:2628363.2628375} or when to switch the input and output channels of devices. These  two top-level categories and nine sub-categories form the basis of design questions posed in our expert survey described in Section~\ref{section:expert-survey}.

\subsection{Technological Challenges}
There are a number of technological challenges for realizing mobile multi-device ecosystems, including binding, security, spatial registration, heterogeneous platforms and sensors, non-touch interaction as well as development and runtime environments. Twenty-seven papers were classified into this category.


\emph{Heterogeneity} of software platforms (e.g., Android, iOS, Windows Mobile), hardware (e.g., sensors), form factors (e.g., smartwatch, smartphone, smartglass, pico projector) or development environments increases as compared to stationary multi-display systems. Specifically, the heterogeneity of platforms can lead to data fragmentation, which impedes sharing of information between devices \cite{santosa2013field}. 

\emph{Development toolkits} targeting cross-device applications involving mobile devices (e.g., \cite{houben2015watch, yang2014panelrama, nebeling2014xdstudio, chi2015weave, Schreiner:2015:CFC:2702613.2732909}) are proliferating as device heterogeneity increases. To address this, they can, for example, support the distribution of web-based user interfaces across displays with varying characteristics (such as size, distance, resolution) \cite{yang2014panelrama}, allow for on-device authoring \cite{nebeling2014xdstudio} or the integration of hardware sensor modules \cite{houben2015watch}. 

Still, these toolkits have a number of challenges to address in the future. For example, we need better support for creating \emph{user interface widgets} that can adopt themselves to the manifold input and output configurations or awareness~\cite{Garrido:2014:AAU:2598153.2598160} in mobile multi-device environments. Specifically, it remains unclear if existing adaption strategies  (e.g., from responsive web design \cite{gardner2011responsive}) remain valid when users relocate widgets frequently between displays or how they should be operated and appear when spanning across multiple displays (including non-touch displays such as smartglasses) \cite{grubert2015multifi}. 

Also, most existing toolkits have not anticipated the integration of \emph{non-touch screen devices}. More specifically, projection based systems, such as optical see-through head mounted displays, or wearable pico-projectors still need  better integration. 

\emph{Device-binding}, i.e. the association and management of multiple devices into a common communication infrastructure needs to be better addressed for mobile multi-device scenarios. There is a large body of work on technical and user-centred aspects on this topic \cite{chong2014survey} ranging from individual \cite{Chong:2011:UAW:1978942.1979219} to group binding \cite{Chong:2013:GUA:2470654.2466207, Jokela:2013:CET:2470654.2466459, Jokela:2014:FBM:2628363.2628376, Jokela:2015:CDC:2797212.2776887}. Existing techniques are generally not found outside of laboratory contexts. Furthermore, most research has concentrated on binding of stationary systems or mobile touch-screen devices such as smartphones and tablets \cite{Jokela:2015:CDC:2797212.2776887}, neglecting the diversified input and output space of new devices such as smartglasses, or wearable activity trackers and smartwatches.

\emph{Security aspects} of mobile multi-device environments have not been a core focus of existing research, with only some exceptions, e.g., regarding second screen apps for ATMs \cite{Regal:2013:MMW:2493190.2493211} or security in group binding \cite{kainda2010two}. 

\emph{Mobile and unified sensing} is another important challenge for creating mobile multi-device systems. So far, we see a fragmented input space for operating individual devices. For example, smartphones and smartwatches typically allow for touch input on their interactive surface or distance sensing with computer vision~\cite{Kristensson2012a} or other sensors. Commercially available smartglasses often use indirect input via a touch pad. Sensing around individual devices has also been explored allowing above surface input on phones (e.g., Project Soli) and smartwatches \cite{houben2015watch} or mid-air input in front of smartglasses (e.g., Microsoft Hololens). Gestures using the devices themselves can also be realized, e.g., through inertial sensors or linear accelerometers. Some mobile phone (e.g., the Nokia N900) posses multiple atenna which can be used for sensing the relative position of other devices \cite{belloni2009multi} and which have been employed for multi-device, collocated interaction \cite{Lucero:2011:PCU:1978942.1979201,Lucero:2012:MCU:2371574.2371634}
However, it remains unclear how to utilize these diverse sensing approaches to create a unified and seamless interaction space across devices. Also, tracking the full six degrees of freedom poses, from all multiple wearable devices, hence enabling a precise mutual spatial understanding of the display positions in space, has been not extensively explored in mobile scenarios and is so far often restricted to lab-based prototypes. For example, approaches such as MultiFi \cite{grubert2015multifi} or HuddleLamp \cite{radle2014huddlelamp} typically rely on stationary tracking systems. Only recently, we see the emergence of mobile sensing solutions, which so far are either restricted in the achievable degrees of freedom or the accuracy and precision of sensing \cite{Jin:2015:TAM:2807442.2807475, Jin:2015:CPA:2807442.2807485}. Similarly, when using head-mounted displays in a spatially registered multi-device environment, we need better and more robust means for calibrating them relative to the user's eye \cite{grubert2010comparative, itoh2014interaction}.

Further challenges include \emph{authoring} mobile multi-device interactions, e.g., for non-experts, in-situ on mobile devices or creating body-referenced information spaces, which ``float’’ virtually around the users' body instead of coinciding with a physical screen \cite{wagner2013body, ens2014personal, grubert2015multifi}. Similarly, the specification of spatial gestures for triggering actions (e.g., through programming by example) has not been studied in this context. Finally, performance issues for web-based frameworks are still a hurdle to allow for fluid interaction across computationally restricted wearable devices \cite{chi2015fw}. 

These eight sub-categories form the basis of technical questions posed in our expert survey described in Section~\ref{section:expert-survey}.

\subsection{Social Challenges}
New technologies and design can lead to new social challenges. While existing social challenges can help inform the design and development of technologies. Considering these as socio-technical systems can help better position the social challenges as considerations to be addressed throughout, rather than simply before or after any technical or design decisions are made. As such, we present five key and durable social challenges that mobile multi-device ecosystems present, including privacy, social acceptability, social participation, social exclusion and social engagement. Four papers in the domain of multi-device environments involved social challenges.

\emph{Privacy} presents a major challenge in the use of public or semi-public displays as part of a mobile multi-device ecosystem~\cite{huang2003semi}. We can consider such forms of social interaction with technology at different scales from inch (cm) to chain (several m) and beyond~\cite{Terrenghi:2009}. Personal devices overcome the privacy challenge by use of private environments, use at an intimate distance, privacy screens or non-visual modalities. Questions arise when we consider how we might share content on intimate displays \cite{Pearson:2015:TSP:2702123.2702247, Kleinman:2015:EMD:2785830.2785833}, at varying scales, different social interaction types or even share content spanning multiple private displays.  For example, users might be reluctant to surrender the possession of their smartphone in group binding situations \cite{Uzun:2011:PDS:1978942.1979282}. We can differentiate between \emph{personal and public privacy}. Personal privacy describes the challenges faced when using personal display elements in a mobile multi-device environment. Public privacy describes the challenges faced when using semi-public and public display elements in a mobile multi-device environment. 

\emph{Social acceptability.} The use of wearable on body displays presents a range of social acceptability issues. Some of the inherent form factors can present acceptability challenges. In addition, existing research has explored the suitability of different parts of the body for gestural inputs~\cite{harrison2014implications}, along with issues of social norms and behaviour~\cite{Profita:2013:DMM:2493988.2494331}. Here, mobile multi-device environments introduce new challenges as the coordination and movement of multiple displays can require unusual inter-display coordination and body orientation. Also, in contrast to touch-only operated displays such as smartphones, the manipulations of multiple body proximate displays through spatial gestures are more visible whereas the effects of those actions remain hidden to bystanders~\cite{reeves2005designing}. Depending on the social situation this could lead to inhibited or non-use of an interactive system, similar to observations made for handheld Augmented Reality systems \cite{grubert2012playing, grubert2013playing}. Further issues arise from the use of shared or public display elements within an ecosystem~\cite{huang2003semi}. All of these issues are modulated by differences in cultures, work practices, age, familiarity with technology an evolving social norms for new technology behaviours.   

\emph{Social participation.} Today, civic discourse is impacted by the isolation that technologies provide people. For example, the ``filter bubble''~\cite{Pariser:2011:FBI:2029079} stems from the personalisation in search results presented to individual people. Such bubbles can socially isolate people from one another, into their own economic, political, cultural and hence ideological groups. With mobile multi-device ecosystems, we might further encourage people into ``interaction bubbles'' which isolate them further from others and discourages interpersonal interaction. The ``in-your-face nature'' of what is proposed in mobile multi-display ecosystems, is unlike other forms of technology. One approach to overcome participation is to design technologies to entice users to participate~\cite{brignull2003enticing}.

\emph{Social exclusion.} Mirroring the problems in social participation are the further challenges of social exclusion~\cite{byrne2005social}. By augmenting our interactions with mobile multi-device ecosystems we are changing the nature of our interaction with the world. Many personal technologies reside out of sight, whereas wearable and on body displays present a visible digital alienation to those without access to such technology. By allowing some to see and experience more than others can see are we further disenfranchising people? Do these technologies exacerbate the digital social stratification we are already witnessing? 

\emph{Social engagement.} In using semipublic or public displays as part of an egocentric mobile multi-device ecosystem, issues of performance and social engagement present themselves~\cite{muller2010requirements}. These challenges are also opportunities for improved social engagement between people but also draw into question the appropriateness of any device appropriation. Fair use, sharing space or time, along with the use of non-visual modalities present challenges for the design and deployment of such systems.    

Further challenges include \emph{personal space}, which describes the physical space immediately surrounding someone~\cite{hall1969hidden}, into which encroachment can feel threatening or uncomfortable as well as \emph{fair sharing}, which describes the equitable and joint use of display resources and space.  These five categories form the basis of technical questions posed in our expert survey described in Section~\ref{section:expert-survey}.

\subsection{Perceptual and Physiological Challenges}
There are a number of Perceptual and Physiological challenges for realizing mobile multi-device ecosystems when we consider human perception in mobile multi-device ecosystems from physiological to cognitive levels. Such issues stem from varying display resolutions, luminance, effective visual fidelities, visual interference, color or contrast in display overlap which can be experienced with body proximate ecosystems. Thirteen papers were associated with this category.






\emph{Display switching.} Existing research has identified the cost of display switching~\cite{rashid2012cost} and the factors which influence visual attention in multi-display user interfaces~\cite{rashid2012factors, Cauchard:2011:VSM:2047196.2047256}, specifically for second-screen TV experiences~\cite{Vatavu:2014:VAM:2602299.2602305, Neate:2015:MAS:2702123.2702278, Neate:2015:DAM:2783446.2783613, Brown:2014:HOM:2559206.2578869, holmes2012visual}. These factors include: 
\begin{itemize}
\item \emph{selective attention} \cite{perry1999attention}: the ability to react to certain stimuli selectively when several occur simultaneously.
\item \emph{sustained attention} \cite{perry1999attention}: the ability to direct and focus cognitive activity on specific stimuli.
\item \emph{divided attention} \cite{Brown:2014:HOM:2559206.2578869}: the ability to time-share attention across stimuli; this occurs when we are required to perform two (or more) tasks at the same time and attention is required for the performance of both (all) the tasks.
\item \emph{angular coverage} \cite{rashid2012factors, Cauchard:2011:VSM:2047196.2047256}: the angular extent of the displays in the environment. It can be used to determine if turning one’s body, head of eyes is sufficient for looking at a display.
\item \emph{display contiguity} \cite{rashid2012factors, Cauchard:2011:VSM:2047196.2047256}: the extent to which the proximity or overlap of displays causes them to be associated as continuous or discontinuous.
\item \emph{time to switch between displays} \cite{rashid2012cost, Neate:2015:MAS:2702123.2702278} : describes the time taken to switch one’s gaze from one display to another. This may be due to a combination of eye, head and body movements but does not include time to focus the eyes due to any depth disparity.    
\item \emph{content coordination} \cite{rashid2012factors}: refers to how the content of different displays are semantically connected even when showing different views of the same data. Existing methods have explored cloned, extended and coordinated displays. 
\item \emph{input directness} \cite{rashid2012factors}: refers to the traditional HCI categorisations of input in terms of direct manipulation can be considered as direct, indirect or hybrid. Measures of directeness could aid in understanding physical challenges in such systems. 
\item \emph{input-display correspondence} can be considered as local, global or redirected in mobile multi-device ecosystems.
\item \emph{visual overload} \cite{Neate:2015:MAS:2702123.2702278, Neate:2015:DAM:2783446.2783613}: the over stimulation of the visual sensory system due to outputs from the multi-device environment coupled with the physical environment which can be mitigated with techniques which are aware of where a person is looking~\cite{Dostal2013}.

\end{itemize}




\emph{Focus in human vision.} The shape of our lens and iris alters how much light enters and how our eyes focus. However, our eyes cannot focus sharply on two displays which are near and far simultaneously. If the virtual display plane of an optical see-through head-mounted device is in sharp focus, then effectively closer or distant displays won't be. \emph{Depth disparity} describes a display environment where one’s eyes are regularly changing focus. This occurs when the eye to display distances vary such that the eye is constantly accommodating between display switches. This can be easily seen with a smartwatch which is in focus but is then surrounded by unfocused input from displays effectively further from the eye. The effective distance, not actual distance, needs to be considered as devices, such as optical see-through displays (e.g., Google Glass) often employ optical techniques to generate images at distances which are easier for the eye to accommodate.  A further issue to consider is that as the ciliary muscles in our eyes age, our range of accommodation declines. 
 Another byproduct of our eyes inability to focus sharply on two distances, is that it then takes time for the eye to refocus on objects at different distances. In addition, the speed of this process also declines as the muscles age.  However, with mobile multi-device ecosystems the eye will need noticeable amounts of time (e.g., 300~msec latency and 1000~msec stabilisation period~\cite{charman-2008}) for the focal power of the eye to adapt in markedly discontiguous display spaces. Further, these accomodation times don't include movements if the displays are ``visually field discontiguous''~\cite{rashid2012factors}.

\emph{Field of view} Humans have a limited field of view and an age diminished ``useful field of view'' (UFOV)~\cite{Richards-2006}, which needs to be considered. Excluding head rotation, the typical field of view for a human has a difference between the horizontal and vertical field of view, an area of binocular overlap and areas of monocular far peripheral vision. ``For many of our interaction tasks the UFOV varies between younger and older people. A 36 degree field of view will be practical in many situations'' ~\cite{Richards-2006}. Within each person's field of view we can also distinguish regions of central (ie. foveal, central, paracentral and macular) and peripheral (near, mid and far) vision. The useful field of view, typically includes both central vision, measured through \emph{visual acuity} (ability to distinguish details and shapes of objects), and largely near \emph{peripheral parts of vision} (part of vision that occurs outside the very center of gaze).

Further factors include \emph{change blindness} \cite{simons1997change, Neate:2015:MAS:2702123.2702278} (the phenomena of a change in the visual stimulus (eg. a new icon~\cite{Davies:2012:CMI:2208516.2208606}) being introduced but the observer not noticing it, specifically the introduction of an \emph{obvious change}; it can occur when the stimulus changes slowly or the stimulus is interrupted, for example with a blank display, blink or saccade). By contrast, \emph{inattentional blindness} \cite{mack1998inattentional} (the phenomena of an \emph{unexpected} visual stimulus not being noticed as one’s attention is engaged on other aspects of the visual scene) and visual discomfort (symptoms of visual fatigue or visual distortion)~\cite{Simons:1999fk}.

\section{Expert Survey}
\label{section:expert-survey} 

The goal of the expert survey was two-fold. First, we wanted to complement the literature research to saturate the list of factors we previously identified. Second, we wanted to find out if certain factors were assessed as more important than others by a majority of experts in the field. 

\subsection{Design and Procedure}
The survey was targeted at experts in mobile multi-device interaction or related fields. Experts  were invited through personal e-mail communication. In addition, social media channels were used to reach out to further experts in the field. The main part of the survey consisted of four sections: development, design, social and perceptual/physiological challenges. Participants were free to skip individual sections. In each section, participants were asked to rank a list of factors according to how important they assessed this factor. Furthermore, participants were asked to list any additional factor, which was not included in our list. The survey took about 5-30 minutes to complete, depending on the number of sections participants were willing to answer. One Amazon voucher worth 30 Euros was raffled among participants.

\subsection{Participants}
\begin{figure}[t]\center
\includegraphics[width=0.7\textwidth]{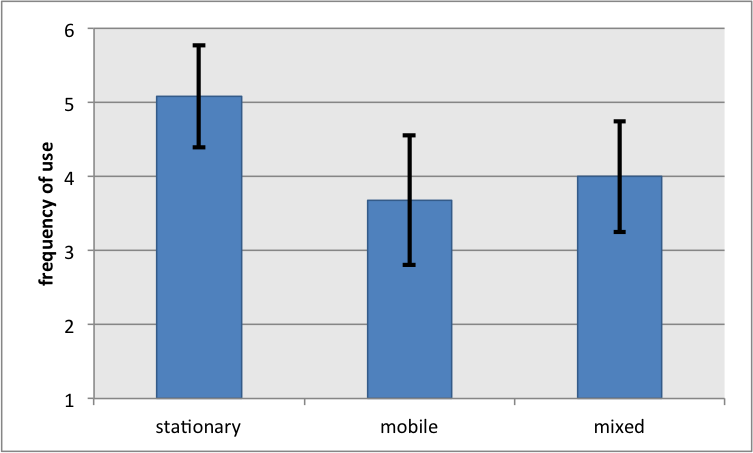}
    \caption{Usage frequency of multi-display environments on a six-item Likert item scale (1: never 6: very frequently). Legend: \emph{stationary}: use of multiple stationary displays (including a notebook + additional external display), \emph{mobile}: multiple mobile displays (e.g., work across a smartphone and tablet or smartphone and smartwatch), \emph{mixed}: mixed mobile and stationary displays (e.g., second screen apps for TVs).}
    \label{fig:mduse}       
\end{figure}

Twenty-seven volunteers participated in the survey (24 male, 2 female, one preferred not to indicate the gender, mean age 33.4 years, SD=6.3). Nineteen participants had experience in designing, developing or evaluating multi-device environments, 20 indicated to have undertaken general research in this area and one participant indicated to teach in this domain. Most participants regularly used stationary multi-display environments, but to a lesser extent mobile and mixed environments, see Figure \ref{fig:mduse}.

\subsection{Results}

We present results for the individual sections on design, development, social and perceptual/physiological challenges next.

\subsubsection{Design Challenges}
\begin{figure}[t]\center
\includegraphics[width=0.7\textwidth]{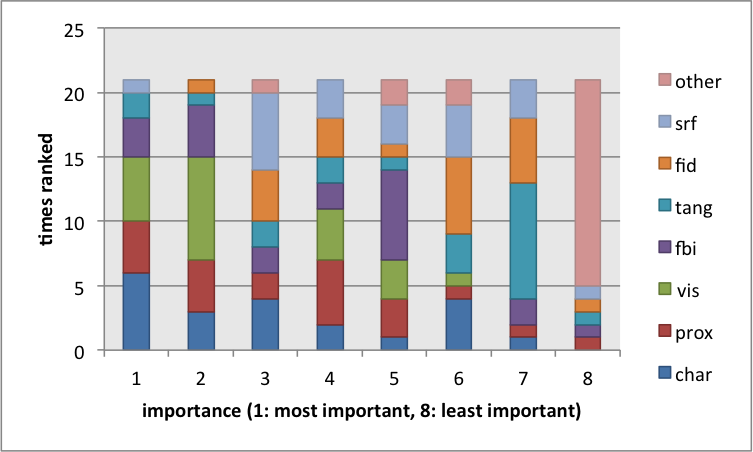}
    \caption{Rankings of design challenges. Legend: \emph{char}: Characteristics of individual devices, \emph{srf}: Spatial reference frame, \emph{fbi}: Foreground-background interaction, \emph{prox}: Proxemic dimensions, \emph{vis}: Visibility, \emph{tang}: Tangibility, \emph{fid}: Fidelity gaps}
    \label{fig:designc}       
\end{figure}

Twenty-one participants answered the design challenges section. 
Figure \ref{fig:designc} shows the ranking on how important individual development factors were assessed by participants. Figure \ref{fig:designca} depicts an aggregated version with multiple summed ranks. Characteristics of individual devices, visibility of devices and proxemic dimensions were identified as important factors.  Foreground-background interaction and spatial reference frame tended to be ranked as medium important followed by fidelity gaps, tangibility and other factors.

In addition, participants were asked if they think that there is a sufficient number of design factors to guide the creation of mobile multi-device systems. On a 5-item item Likert scale (strongly disagree ... strongly agree) the average score was 2.76 (SD=1.22), indicating no general trend. Twelve participants strongly disagreed or disagreed, seven agreed or strongly agreed, two were neutral. 

One participant explicitly mentioned user interfaces adaption to different form factors and one conflict resolution in multi-user scenarios.


\begin{figure}[t]\center
\includegraphics[width=0.7\textwidth]{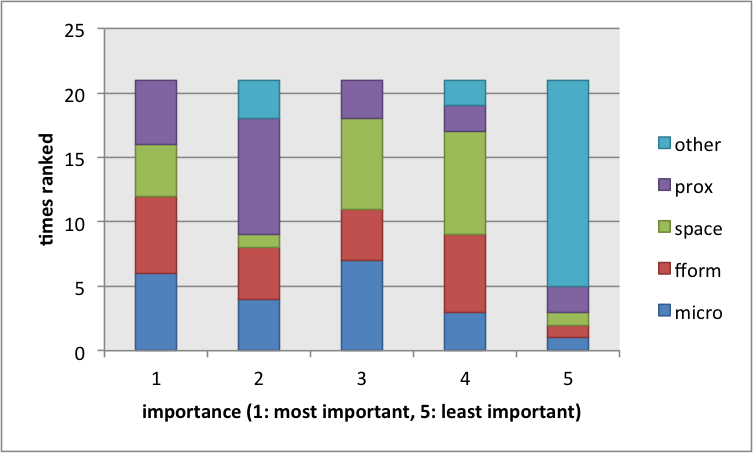}
    \caption{Rankings of design challenges for co-located interaction. Legend: \emph{micro}: micro-mobility, \emph{fform}: F-formations, \emph{space}: Space Syntax, \emph{prox}: Proxemic dimensions}
    \label{fig:designmuc}       
\end{figure}

In addition, participants were asked to prioritize factors for designing multi-device systems for co-located interaction. The results are depicted in Figure \ref{fig:designmuc}. Figure \ref{fig:designmuca} depicts an aggregated version with multiple summed ranks. While no strong trends could be identified, micro-mobility \cite{luff1998mobility, marquardt2012cross} and proxemic dimensions \cite{hall1969hidden, greenberg2011proxemic, marquardt2015proxemic} were ranked as important, followed by F-formations \cite{kendon1990conducting, marquardt2012cross, Marshall:2011:UFA:1958824.1958893} and space syntax \cite{hillier2007space, schroder2007quantifying}. One participant explicitly highlighted accessibility issues (e.g., visibility, reach) when multiple persons interact with distributed multi-device systems.

\subsubsection{Development Challenges}
\begin{figure}[t]\center
\includegraphics[width=0.7\textwidth]{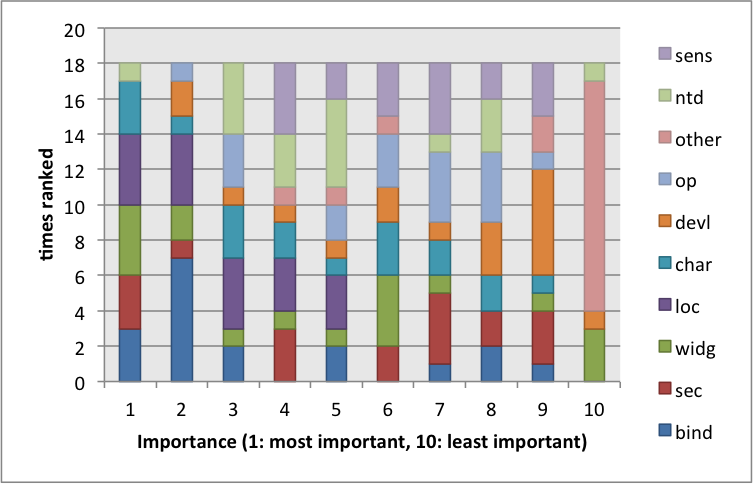}
    \caption{Rankings of development challenges. Legend: \emph{bind}: Ad-hoc binding / joining / leaving device groups, \emph{sec}: Secure communication between devices, \emph{widg}: User interface widget adoption, \emph{loc}: Localization / spatial registration of devices, \emph{char}: Characteristics of individual devices (e.g., contrast, input, output modalities, input output resolution), \emph{devl}: Heterogeneity of development languages, \emph{op}: Heterogeneity of operating systems, \emph{ntd}: Integration of non-touch screen devices (e.g., Google Glass, Microsoft HoloLens), \emph{sens}: Heterogeneity of sensors }
    \label{fig:devc}       
\end{figure}

Eighteen participants answered the development challenges section.
Figure \ref{fig:devc} depicts the ranking on how important individual development factors were assessed by participants. Figure \ref{fig:devca} depicts an aggregated version with multiple summed ranks.
Ad-hoc binding, localization / spatial registration of devices and security were ranked as very important. Integration of non-touch screen devices, characteristics of individual devices, heterogeneity of operating systems tended to get assigned medium priorities. Heterogeneity of development languages, heterogeneity of sensors, UI widget adoption tended to be ranked as medium to less important, but with a wide spread.

One participant mentioned responsiveness and reliability of network-based operations and two testing and debugging, with one highlighting the need for a better support for non-expert developers and ``lack of development support on mobile devices''.


\subsubsection{Social Challenges}
\begin{figure}[t]\center
\includegraphics[width=0.7\textwidth]{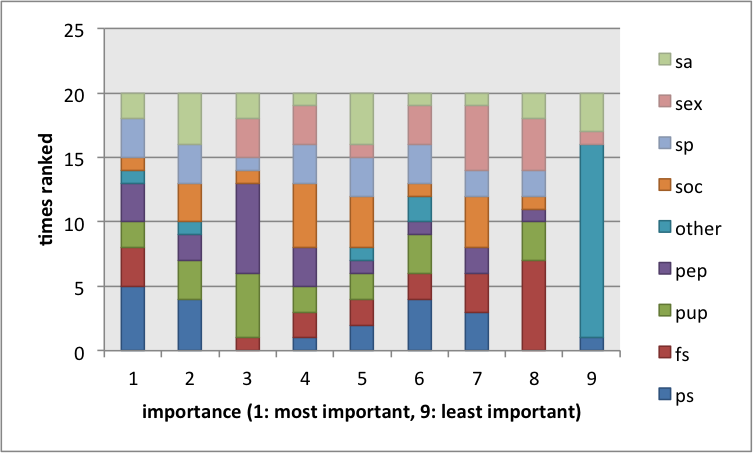}
 \caption{Rankings of social challenges. Legend: \emph{ps}: personal space, \emph{fs}: fair sharing, \emph{pup}: public privacy, \emph{pep}: personal privacy, \emph{soc}: social engagement, \emph{sp}: social participation, \emph{sex}: social exclusion, \emph{sa}: social acceptability}
    \label{fig:socc}       
\end{figure}

Twenty participants answered the social challenges section. Figure \ref{fig:socc} indicates the ranking on how important individual social factors were assessed by participants. Figure \ref{fig:socca} depicts an aggregated version with multiple summed ranks. The social factors were ranked diversely, not indicating a strong trend for most factors. However, social exclusion and fair sharing tend to be ranked as less important. One participant suggested that for social participation one should understand more the joint participation or co-interaction of multiple users instead on focusing on isolation aspects. Another participant mentioned social exclusion due to platform differences.

\subsubsection{Perceptual and Physiological Challenges}
\begin{figure}[t]\center
\includegraphics[width=0.7\textwidth]{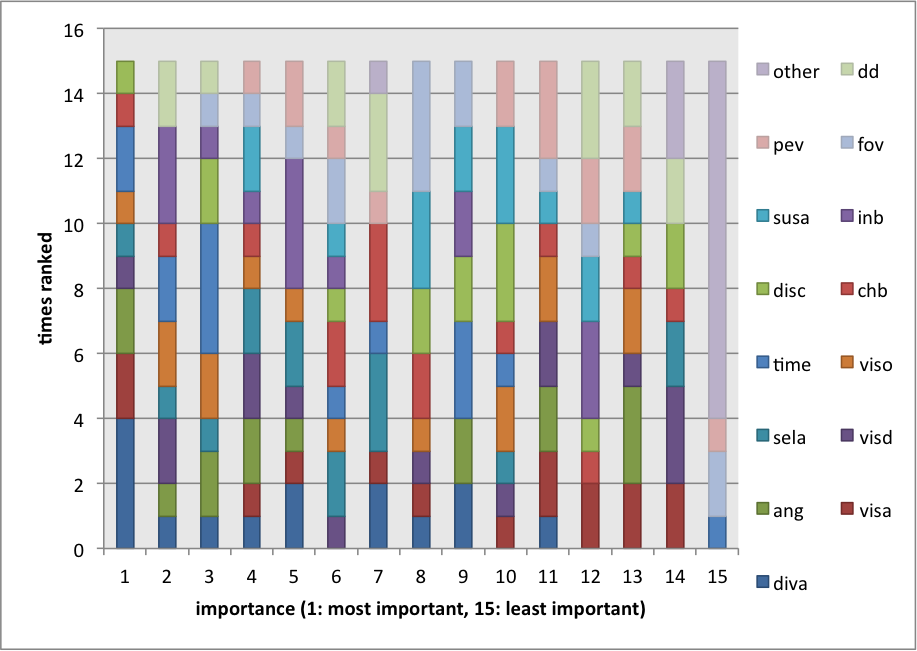}
 \caption{Rankings of perceptual and physiological challenges. Legend: \emph{diva}: divided attention, \emph{ang}: angular coverage, \emph{sela}: selective attention, \emph{time}: time to switch between displays/devices, \emph{disc}: display contiguity, \emph{susa}: sustained attention, \emph{pev}: peripheral vision, \emph{visa}: visual acuity, \emph{visd}: visual discomfort, \emph{viso}: visual overload, \emph{chb}: change blindness, \emph{inb}: inattention blindness, \emph{fov}: field of view}
    \label{fig:perc}       
\end{figure}

Fifteen participants answered the section on perceptual and physiological challenges. Figure \ref{fig:perc} depicts the ranking on how important individual factors were assessed by participants.  Figure \ref{fig:perca} depicts an aggregated version with multiple summed ranks. The factors were ranked diversely, not indicating a strong trend for most factors. However, divided attention, angular coverage, selective attention, visual overload, visual discomfort, inattention blindness and time to switch between devices were identified as more important. No other factors were mentioned by the participants.

\subsection{Discussion of the Survey Results}
One goal of this survey was to saturate relevant factors for the creation and use of mobile multi-device environments. The experts identified only a few additional factors, including accessibility issues (e.g., visibility, reach) and development support for non-experts and development tools for mobile platforms. This suggests the identified factors can form a basis for future exploration and new research and development in mobile multi-device excosystems emerge. 

Another finding of our survey is, that participants consistently identified only some development challenges (e.g., ad-hoc binding, localization / spatial registration of devices) and design factors (e.g., device characteristics and proxemic dimensions) as important. 
Beyond these selected factors, no strong consensus on the importance of the diverse factors was found. This could indicate that the importance of individual factors is very dependent on the context of use. In fact, one user explicitly mentioned that ``I think the order of importance of these challenges depends on the users, the context and the system under development''. One clear outcome from this survey is the need to establish new theories and research motor themes~\cite{kostakos2015big} for mobile multi-device ecosystems. Without these, research and developments in this area will remain fragmented, diverse and disconnected from any theoretical grounding. 


\section{Discussion}

Through our literature survey and expert survey we have identified a number of challenges for mobile multi-device environments. While some of these challenges are similar to stationary multi-display systems, the highly mobile nature of the components leads to a large number of challenges to be addressed including the need for well-founded theory.  

For design challenges, we see a large number of proposals on what factors and frameworks are relevant for creating mobile multi-device systems. Still, there is no strong consensus in the community on if the existing factors are sufficient to guide the design of current and future systems. Only some design factors (eg. proxemics, visibility, characteristics of individual devices) were consistently identified as important by experts. However, that does not necessarily imply that other factors are less important, but that those factors are either more context-dependent or just not well researched in the community. For example, we believe that with the diversification of input and output channels in mobile multi-device scenarios, we need to incorporate better the relative differences between device capabilities (ie. fidelity gaps), not just their individual absolute characteristics. One such example is the transition between touch and mid-air interaction. While recent research has shown that for some tasks (e.g., gaming) users would prefer mid-air input for smartglasses \cite{tung2015userdefined}, there are clearly benefits of haptic qualities of surfaces \cite{harrison2010appropriated}, which are evident in touch being the dominant interaction mode for smartphones and smartwatches. While researchers have begun to investigate the joint interaction space of touch and free-space input (e.g., \cite{marquardt2011continuous, chen2014air+, kim2015retro}), there is clearly a larger design space to explore in highly mobile multi-device scenarios. Another opportunity might be to further investigate micro-mobility for co-located interaction \cite{luff1998mobility, marquardt2012cross}. The increasing number of mobile and wearable displays open up new possibilities to study how people utilize interactive mobile devices to share or hide information from others \cite{Pearson:2015:TSP:2702123.2702247, Ens:2015:CIR:2807442.2807449}.

Looking at technological and development challenges we see that device-binding is considered as a very important topic. However, so far device-binding has mainly been considered for tablets and smartphones \cite{Jokela:2015:CDC:2797212.2776887}. There is still potential to find novel ways to bind other mobile devices such as smartglasses, smartwatches, pico-projectors or activity-trackers without a display. We also argue, that there is an increased need for considering the adoption of user interface widgets across devices. While there are guidelines how to change the layout of widgets depending on different screen sizes (e.g., from responsive web design \cite{gardner2011responsive}), those guidelines often assume the interaction on an individual device at a time. It remains to be explored how well users can interact with changing layouts if they have to relocate widgets frequently between displays (e.g., a smartwatch and tablet). Also, there is more research needed for how to adopt widgets that span multiple displays at once, including  non-touch displays such as smartglasses. Is it sufficient to change the appearance of a widget to a different level of visual details or do the semantics of operation have to change \cite{grubert2015multifi}? Furthermore, we see the  opportunity to combine device-integrated \cite{Jin:2015:TAM:2807442.2807475, Jin:2015:CPA:2807442.2807485} with body-mounted sensors \cite{chan2015cyclops, chan2015cyclopsring} into hybrid pose tracking systems in order to derive a full spatial understanding of all on-and around the body devices. However, to date it has not been explored in depth how precise and reliable those mobile sensing solutions can and should work \cite{mulloni2012experiences}. Furthermore, it has still to be explored which granularity of spatial sensing (precise to none) is actually sufficient for various cross-device interaction tasks. Finally, as many cross-device toolkits offer to create web-based user interfaces it might be worthwhile to investigate the integration of sensing solutions based on web-standards \cite{oberhofer2012nft}.

Our literature review and survey suggests the consideration of social challenges in mobile multi-device ecosystems is immature. The ecological validity of the scenarios described in many papers are open to criticism due to the unconvincing use cases, novel forms of interaction or unrealistic scenarios described. The laboratory settings can contribute new research findings to many of the other facets described while our social challenges require research in non-technical domains or socio-technical settings. 

Finally, the perceptual, cognitive and physiological issues will clearly play a more important role in studying mobile multi-device environments in the future. However, this research should not remain in a HCI context alone as it requires a wider range of research expertise. An example of this can be seen in investigation of some issues (e.g., attention) in works on interactive TV / second screen experiences ~\cite{Vatavu:2014:VAM:2602299.2602305, Neate:2015:MAS:2702123.2702278, Neate:2015:DAM:2783446.2783613, Brown:2014:HOM:2559206.2578869, holmes2012visual}, but is less studied in more mobile usage scenarios. 

In the future the survey results could be complemented with further studies targeted at end-users of multi-device environments, e.g., similar to the work of Jokela et al. \cite{jokela2015diary}.



\section{Conclusion}
There are many future visions of computing~\cite{Quigley:2013:vvc} which incorporate aspects of mobile multi-device ecosystems. Within this article, we have considered design, technical, social and perceptual challenges and the questions raised in interaction with mobile multi-device environments. The fundamental challenges in mobile multi-device ecosystems reach beyond that of stationary multi-display ecosystems, due to the larger variety of input and output modalities, the mobility of its individual components and due to the proximity of those devices to the human body. We have based our findings both on a literature survey and on an expert survey. While the expert survey indicated that we have identified a large number of current challenges, there is only little agreement on the importance of individual challenges. This might be due to the highly contextual nature of mobile multi-device interaction, which influences the importance of individual factors. By presenting current challenges and questions we hope to contribute to shaping the research agenda for new theory, new areas of research inquiry outside of HCI and research on the interaction with mobile multi-device environments.

\begin{backmatter}


\section*{Author's contributions}
JG carried out the digital library search and drafted the design and technological challenges. AQ drafted the social, perceptual and physiological challenges. JG, MK and AQ conceived of the expert survey, and participated in its design and coordination and all helped to draft the manuscript. All authors read and approved the final manuscript.


\bibliographystyle{bmc-mathphys} 
\bibliography{mux}      

\section{Appendix}

\begin{figure}[t]\center
\subfloat{\includegraphics[width=0.7\textwidth]{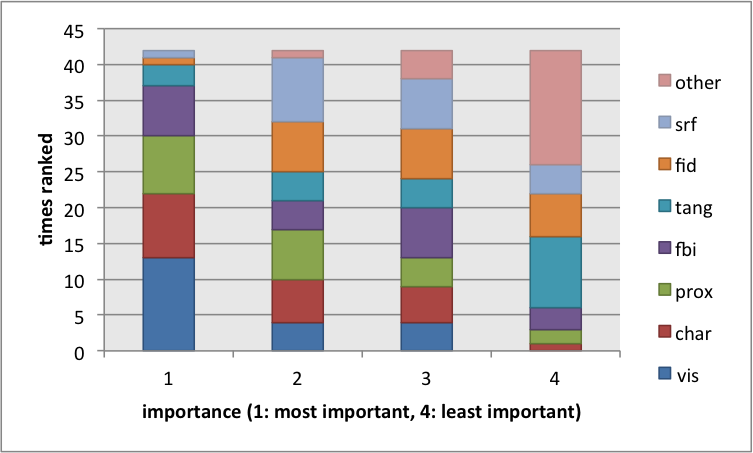}}~\\
\subfloat{\includegraphics[width=0.7\textwidth]{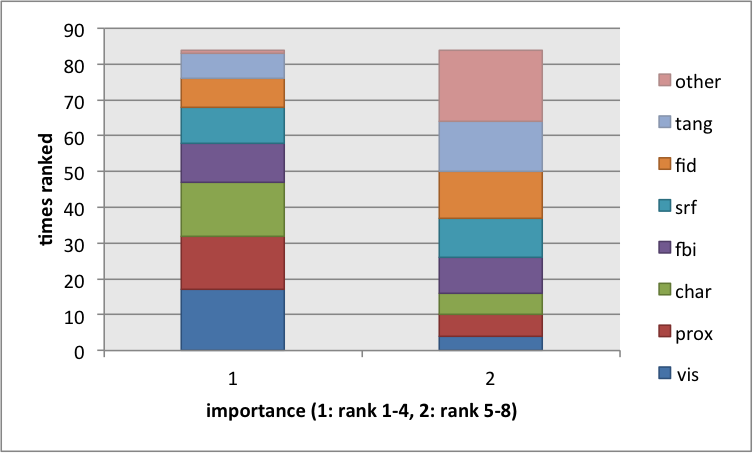}}
    \caption{Aggregated rankings of design challenges. Legend: \emph{char}: Characteristics of individual devices, \emph{srf}: Spatial reference frame, \emph{fbi}: Foreground-background interaction, \emph{prox}: Proxemic dimensions, \emph{vis}: Visibility, \emph{tang}: Tangibility, \emph{fid}: Fidelity gaps}
    \label{fig:designca}       
\end{figure}

\begin{figure}[t]\center
\subfloat{\includegraphics[width=0.7\textwidth]{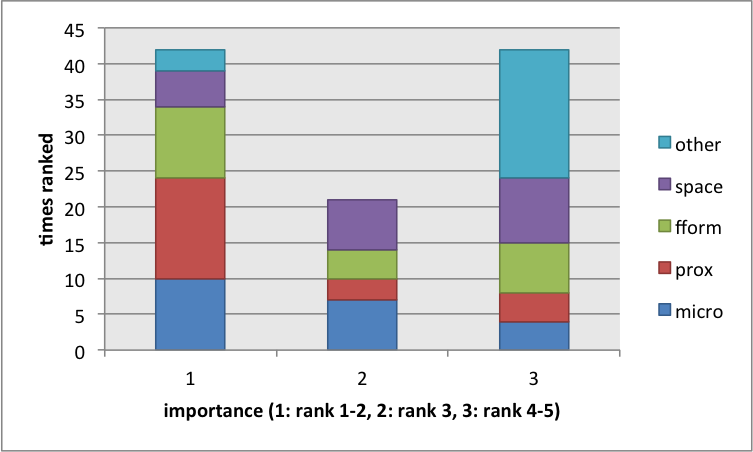}}
    \caption{Aggregated rankings of design challenges for co-located interaction. Legend: \emph{micro}: micro-mobility, \emph{fform}: F-formations, \emph{space}: Space Syntax, \emph{prox}: Proxemic dimensions}
    \label{fig:designmuca}       
\end{figure}

\begin{figure}[t]\center
\subfloat{\includegraphics[width=0.7\textwidth]{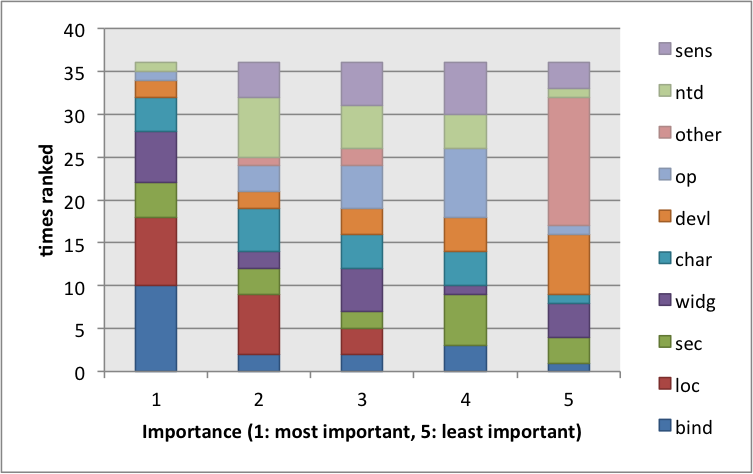}}~\\
\subfloat{\includegraphics[width=0.7\textwidth]{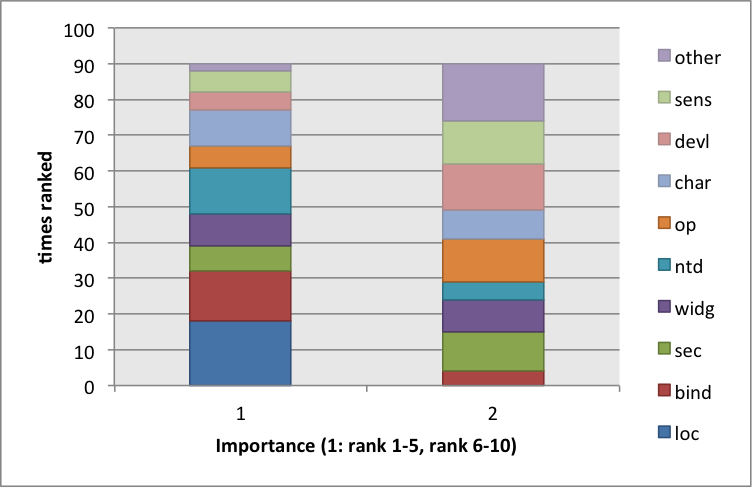}}
    \caption{Aggregated rankings  of development challenges. Legend: \emph{bind}: Ad-hoc binding / joining / leaving device groups, \emph{sec}: Secure communication between devices, \emph{widg}: User interface widget adoption, \emph{loc}: Localization / spatial registration of devices, \emph{char}: Characteristics of individual devices (e.g., contrast, input, output modalities, input output resolution), \emph{devl}: Heterogeneity of development languages, \emph{op}: Heterogeneity of operating systems, \emph{ntd}: Integration of non-touch screen devices (e.g., Google Glass, Microsoft HoloLens), \emph{sens}: Heterogeneity of sensors }
    \label{fig:devca}       
\end{figure}

\begin{figure}[t]\center
\subfloat{\includegraphics[width=0.7\textwidth]{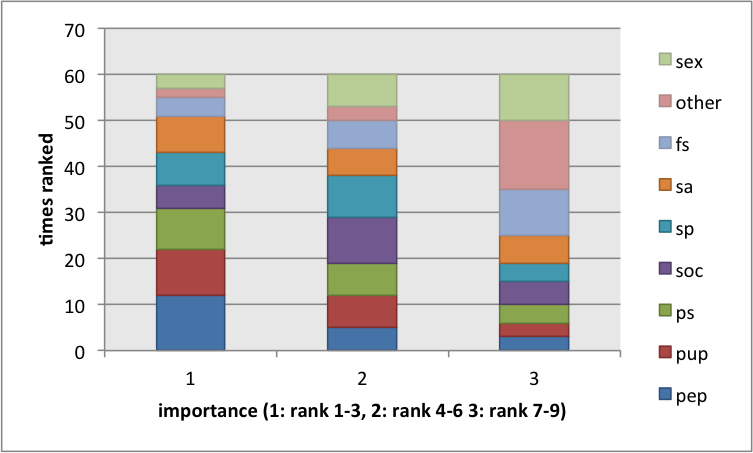}}
 \caption{Aggregated rankings of social challenges. Legend: \emph{ps}: personal space, \emph{fs}: fair sharing, \emph{pup}: public privacy, \emph{pep}: personal privacy, \emph{soc}: social engagement, \emph{sp}: social participation, \emph{sex}: social exclusion, \emph{sa}: social acceptability}
    \label{fig:socca}       
\end{figure}

\begin{figure}[t]\center
\subfloat{\includegraphics[width=0.7\textwidth]{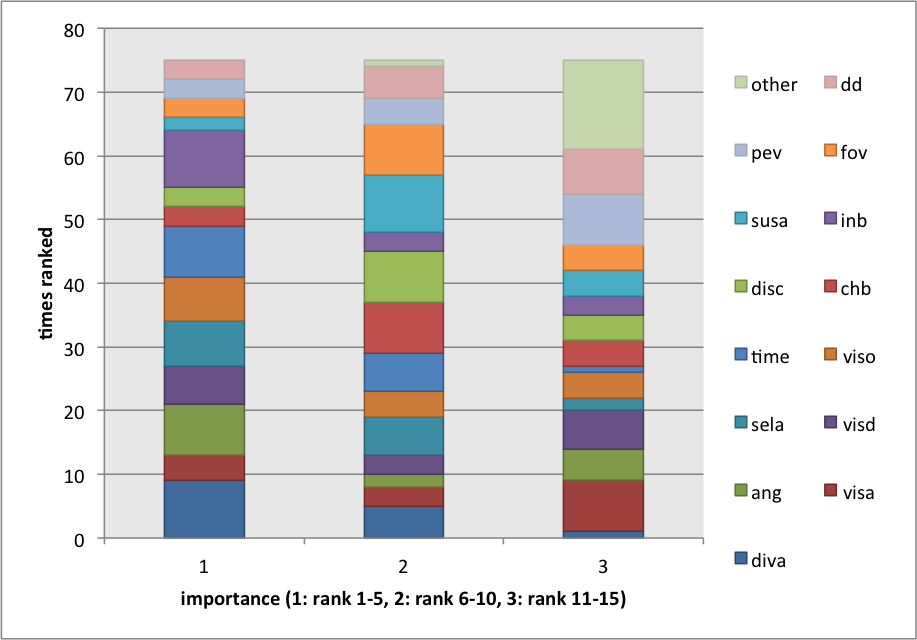}}
 \caption{Aggregated rankings  of perceptual and physiological challenges. Legend: \emph{diva}: divided attention, \emph{ang}: angular coverage, \emph{sela}: selective attention, \emph{time}: time to switch between displays/devices, \emph{disc}: display contiguity, \emph{susa}: sustained attention, \emph{pev}: peripheral vision, \emph{visa}: visual acuity, \emph{visd}: visual discomfort, \emph{viso}: visual overload, \emph{chb}: change blindness, \emph{inb}: inattention blindness, \emph{fov}: field of view}
    \label{fig:perca}       
\end{figure}

\end{backmatter}
\end{document}